# High-Precision Tuning of State for Memristive Devices

# by Adaptable Variation-Tolerant Algorithm

Fabien Alibart[1], Ligang Gao[1], Brian Hoskins[2], and Dmitri Strukov[1]

[1] Department of Electrical and Computer Engineering

[2] Materials Department

University of California Santa Barbara

Santa Barbara, CA 93106, USA

**Abstract**

Using memristive properties common for the titanium dioxide thin film devices, we designed a simple write algorithm to tune device conductance at a specific bias point to 1% relative accuracy (which is roughly equivalent to 7-bit precision) within its dynamic range even in the presence of large variations in switching behavior. The high precision state is nonvolatile and the results are likely to be sustained for nanoscale memristive devices because of the inherent filamentary nature of the resistive switching. The proposed functionality of memristive devices is especially attractive for analog computing with low precision data. As one representative example we demonstrate hybrid circuitry consisting of CMOS summing amplifier and two memristive devices to perform analog multiply and accumulate computation, which is a typical bottleneck operation in information processing.

In order to fully realize analog properties of resistive switching devices [Saw08, Was09, Per11, Kim11, Yan09, Lik08], which are also called "memristive devices" [Chu11], one has to deal with significant variations in the switching behavior. In some applications such variations can be tolerated by the structure of the circuits, such as the case with some versions of artificial neural networks in which memristive devices are implementing synapses [Lik11, Sni07, Yu11b, Seo11, Cha11, Jo10a]. However, for all other applications, e.g., multilevel memory [Bec00, Yu11a, Cho06], configurable filters [Dri10], and analog computing circuits (see, e.g., various theoretical proposals in Refs. [Wei11, Shi11, Per10, Lai10, Pro10]), the performance directly depends on how accurately the resistive state can be set.

A natural way to tackle variations in switching behavior is to utilize active feedback scheme, e.g. applying iterative write and read (test) pulses to converge to certain desired conductive state of the device. Such scheme has been successfully applied to phase change memories to achieve multilevel memory operation [Pap11, Bed09]. A similar idea to use closed-loop circuitry has been proposed and theoretically simulated using Spice model for $TiO_2$ devices [Yi11]. In this paper, we experimentally demonstrate a simple feedback algorithm which takes into account specific memristive behavior of titanium dioxide devices to tune resistance state of the device within 1% relative accuracy within all the dynamic range. We then use our algorithm to demonstrate one of the most important operations in information processing - analog multiply and accumulate (MAC).

The $Pt/TiO_2(30nm)/Pt$ devices have been implemented in "bone-structure" geometry with active area $\sim 1~\mu m^2$ by atomic layer deposition technique (see supplementary information for details). The electrical measurements were carried out with a B1500 Agilent parameter analyzer in ambient conditions.

Figure 1 shows typical resistive switching in titanium dioxide devices. The full loop is obtained by quasi-DC triangular voltage sweep between 0 and -1.5V followed by a quasi-DC triangular current sweep between 0 and 750 µA with a sweeping period of about 4s. Alternatively, the device can be switched continuously between ON and OFF states. The incremental RESET switching is obtained by applying voltage sweeps of increasing amplitude from -0.8V to -1.5V with a step of 10 mV. Similarly, incremental SET transition is obtained with current sweeps of increasing amplitude from 100 to 750 µA with a step of 50 µA.

A more accurate control of the device is possible using sequence of relatively large amplitude write pulses followed by smaller non-disturbing read pulses [Pic09]. Note that for all experiments described below we use voltage-controlled pulses for SET switching also. The main reason is that current-controlled switching (or alternatively the utilization of a compliance transistor) is not compatible for large scale crossbar circuits, even though it could be more natural for SET switching in the context of single devices because it allows avoiding overshooting and overheating. In particular, the measurement is composed of two different sequences of pulses: (i) the read pulses of -200 mV and 1 ms width are used to probe the state of the device which is represented by the resistance or by current measured at -200 mV) and (ii) the write pulses which are used to change the state of the memristive device, whether by changing the pulse width and/or the pulse amplitude. The two pulses (read and write) are alternated at a frequency of 0.5 Hz to prevent the accumulative Joule heating effect from single write pulses.

Figure 2 shows the evolution of the state as a function of both the write pulse amplitude and cumulative time (i. e. summation of the write pulses duration). Before the pulse measurements, the devices are set to a low (high) resistance state with a quasi-DC current (voltage) control loop in order to start the SET (RESET) transition from the same initial condition. In particular, on Figure 2a each curve shows the dynamic evolution of the device's state, as a result of the application of fixed amplitude pulses with an exponentially increasing duration from 200 ns to 1 ms. Figures 2 c and d show in detail the dynamic evolution of the device's state for the most relevant stress conditions (i.e. faster switching) for RESET and SET transition, respectively, when both the amplitude and the width of the write pulses are fixed (a constant pulse width of 200 ns has been used in these measurements) .

Figures 2a, b, c clearly demonstrates that the device can be switched fast - of the order of 1 us for both transitions for maximum applied voltages and retain its state for at least longer than 1 s if the applied voltage is within (-0.6 V < $v$ < 0.5 V) region. Moreover, Figures 2b, c highlights that the switching dynamics is exponential with voltage for SET switching and roughly follows power low for RESET (at least before it saturates) – See supplementary information for more details on fitting. This is similar to previously reported dynamics [Pic09] – though it should be noted that in our case the state is defined as a conductance (or resistance) at specific bias rather than phenomenological parameter such as barrier width. Our experimental results support that the retention (at, say -0.2V) to write ratio (at maximum amplitude) is at least larger than $10^6$ but likely to be much higher. First of all, we did not apply stress long enough to see significant drift of the state so that the retention time is certainly longer. Secondly, exponential nature of SET dynamics is most likely a result of thermal effects [Stra11, Bor09, Shk09, Iel11]. As a result, the switching time could be made much faster if larger write voltages are applied [Iel11, Pic09, Str09]. In our case we avoid checking this hypothesis because experimental setup does not allow applying short pulses, while relative long high-amplitude pulses lead to overstressing and damaging of the device.

While the general dynamics for all our devices after forming is qualitatively similar there are significant variations – both from device to device and for the same device upon cycling, which prevents from using calculated stress voltage to drive the device to the desired state. In principle, such variations can be coped with by overstressing the device when only two extreme resistive states in the device are needed. To tune the device to a specific intermediate state overstressing strategy is clearly not an option and instead we use the feedback algorithm. Our algorithm is based on the fact that large-amplitude pulses can be used to reach desired state faster but also at much cruder precision (Figs. 2 a, b, c) due to the fixed minimum pulse duration (which is limited by experimental setup). Once the device is driven close to the vicinity of the desired state smaller amplitude pulses can be used for fine tuning. (Note that using pulses of relatively small fixed amplitude only is not an attractive option because it may require exponentially long write time.) To take advantage of such switching dynamics our algorithm (Fig. 3) is based on a sequence of increasing amplitude voltage ramps of appropriate polarity (which depends on the initial and the desired state of the devices). In particular, we apply 10 μs-long voltage pulses (starting from 0.5V for the SET sequence and starting from -0.5V to the RESET sequence) with a step of 5mV. Similar to previous measurements the device state is checked with read pulse after each write pulse and the voltage ramp is applied until it reaches and overshoots the desired state. At this point the new voltage ramp of opposite polarity is started. Because this time the initial state is closer to the desired one the maximum amplitude of the voltage pulse in the new ramp (i.e. before the ramp is stopped) is smaller which in turn ensures even better precision. Using this algorithm we were able to tuned the device to 1% precision with respect to the desired state within the whole dynamic range of the device (Fig. 3a).

In principle, even more accurate tuning might be possible by using more ramps (i.e. more time) though in this particular experiment we stopped our algorithm when 1% accuracy was achieved.

Theoretically, i.e. from physical noise analysis, any computation at signal precision below 8-bit (for robotics, distributed sensor networks applications, etc.) could be done much more efficiently in analog or mixed signal circuits [Str07], with wires carrying multiple bits of information [Sar98]. However, even low precision analog processing in conventional CMOS technology is cumbersome, due to the lack of suitable hardware, e.g. to implement multilevel (analog) weights. On the other hand, low precision analog information processing can be implemented with hybrid circuits, which combine single conventional CMOS chip and layer(s) of quasi-memristive devices [Lik08]. Figure 4 illustrates such circuit consisting of discrete chip CMOS summing amplifier and two memristive devices programmed to high precision state with proposed algorithm (See supplementary info). In particular, to realize MAC circuits, the memristive devices implement density-critical configurable low precision weights, while CMOS is used for the summing amplifier, which provides gain and signal restoration. As a result, individual voltages applied to memristive device can be multiplied by the unique weight (conductance) of memristor and summed up by CMOS amplifier - all in analog fashion.

The results demonstrated on Figures 1-4 are for microscale devices, however we strongly believe that it can be sustained in nanoscale devices based on previously reported experimental scaling analysis [Lee11] and the fact that forming process produces nanoscale filaments with active area localized to few nanometers spots [Stra10, Kwo10]. In fact, the nanoscale memristive devices have typically less variations [Jo10b] and, therefore, we expect better precision could be achieved within shorter amount of time using our algorithm.

It is also worth mention that our algorithm was studied in the context of single devices. In the passive crossbar structures semi-selection and leakage currents might present additional challenges. However, at least the former issue should not be a problem for our devices because of the strong nonlinearity in the switching dynamics (Fig. 2a). Also, while there is a natural tradeoff between the tuning time and resulting accuracy one can expect certain physical limit to the precision, e.g. defined by the shot ionic noise and temporal instabilities, which is certainly worthwhile investigating.

**Acknowledgments**

We thank Konstantin Likharev and Ashok Ramu for helpful discussions. This work was supported via NSF grant NSF grant CCF-1028336.

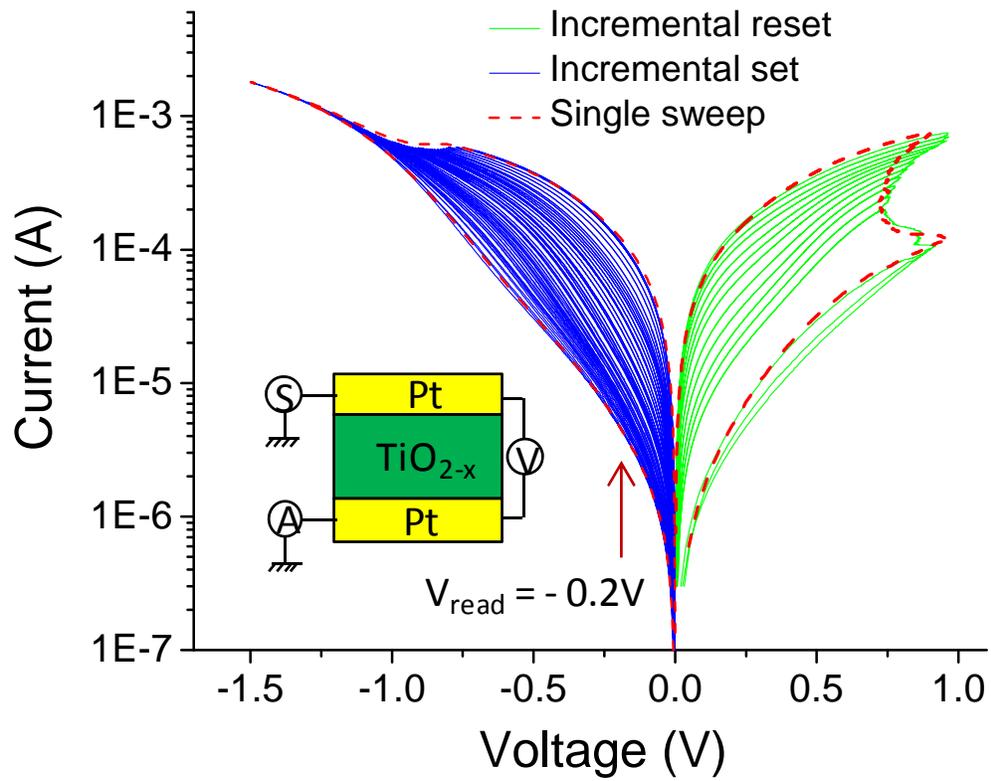

**Figure 1:** Typical *I-V* showing incremental resistive switching of the TiO$_{2-x}$ thin film. The RESET is obtained by a sweep of voltage between 0 and -1.5V and the SET by a sweep of current between 0 and 750uA.

**(a)**

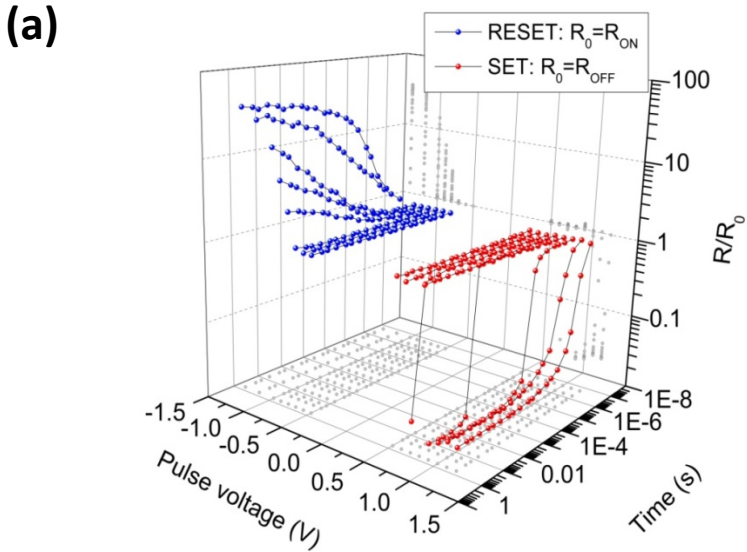

**(b)**

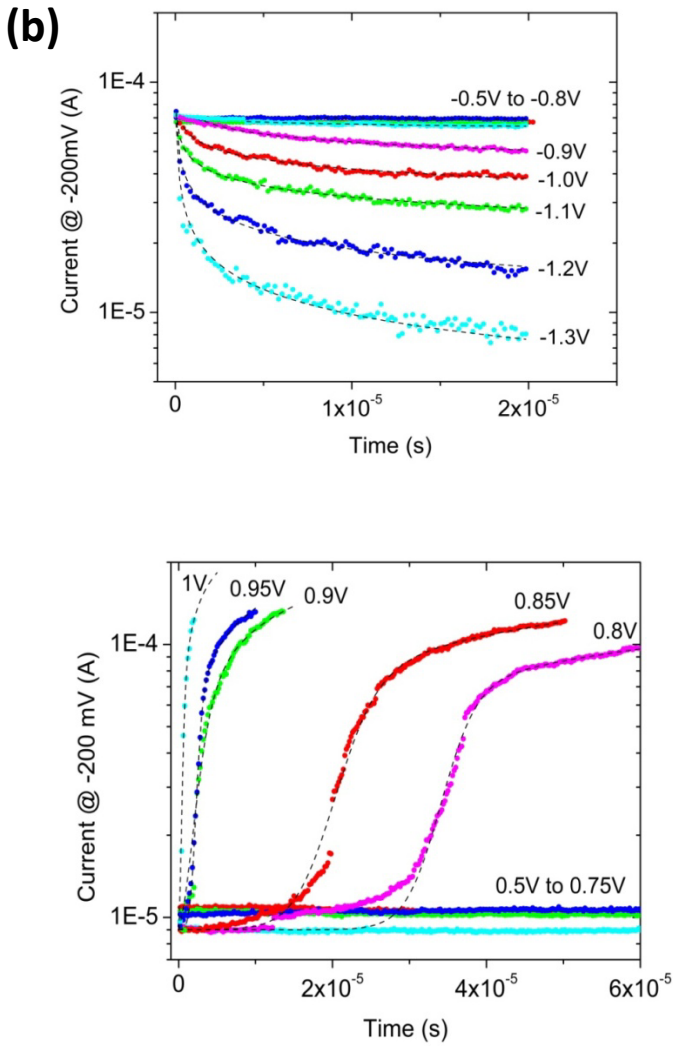

**Figure 2:** Switching dynamics characterized by voltage pulse stress with variable amplitude and duration. Panel (a) shows general switching behavior and volatility characteristics over large duration range. Panels (b) and (c) show detailed transitions over shorter (and more relevant duration range for write operation) and with fitting curves for SET and RESET transitions, correspondingly. Note that before each measurement, the device is set to the same high (low, respectively) resistance state by a voltage (current) sweep.

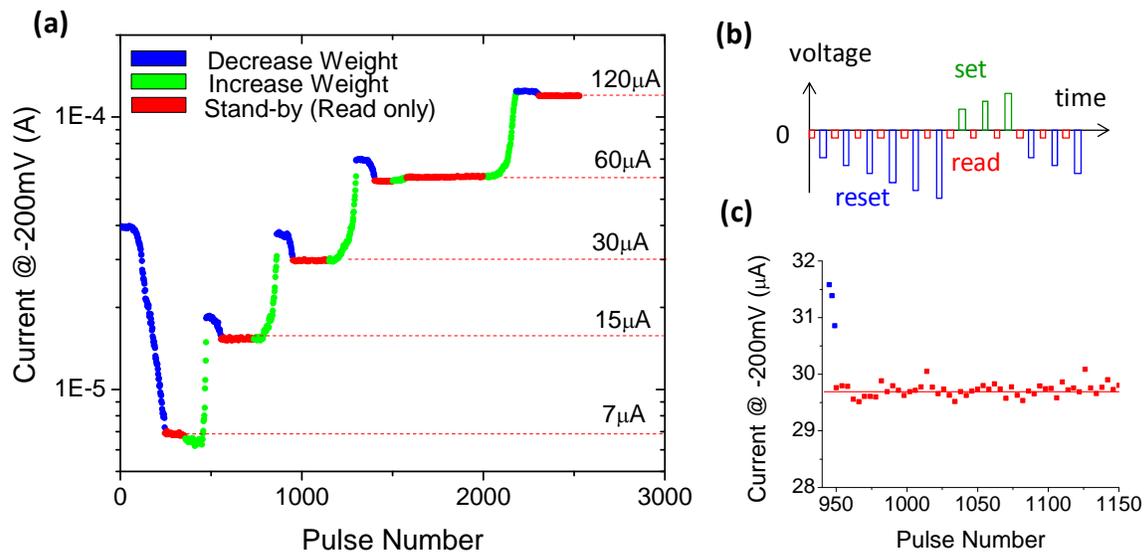

**Figure 3:** Tuning of the device state within 1% accuracy: (a) Demonstration of the algorithm to tune the resistive state of the device to 7μA, 15μA, 30μA, 60μA, 120μA within an accuracy of 1% using the algorithm shown on panel (b). Panel (c) shows zoom-in of the particular intermediate state.

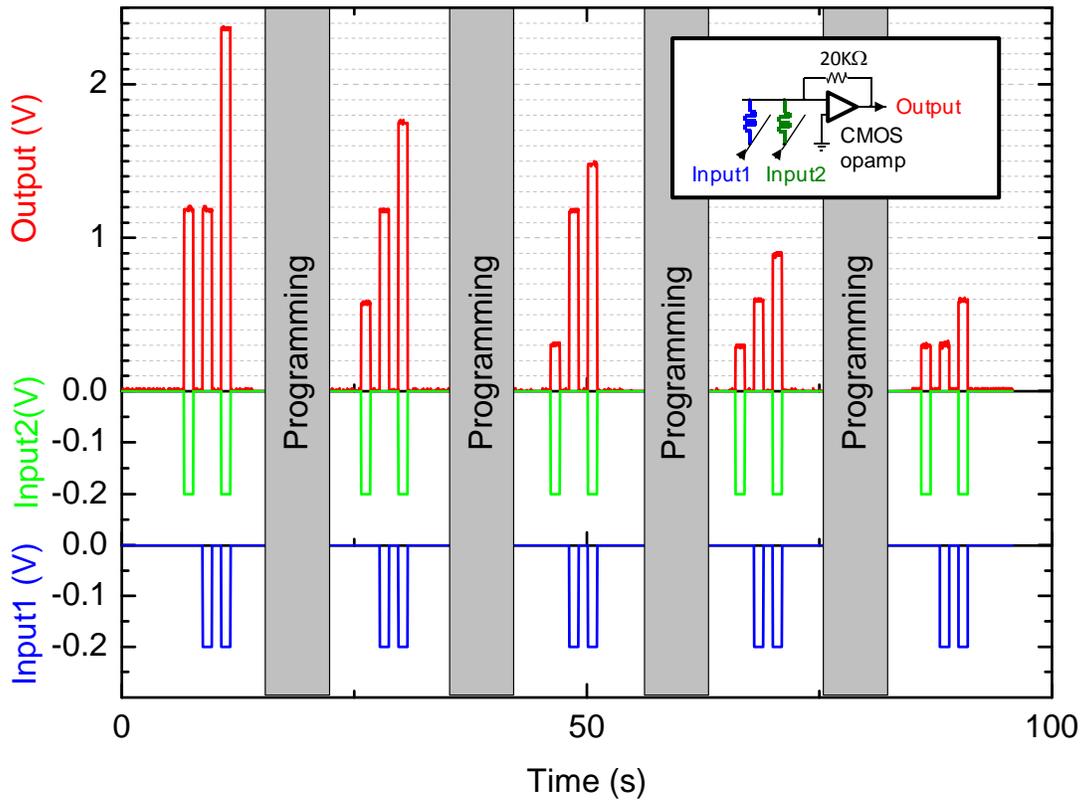

**Figure 4:** Illustration of analog multiply and add circuitry operation with CMOS summing amplifier and memristive devices set with high precision with proposed algorithm.

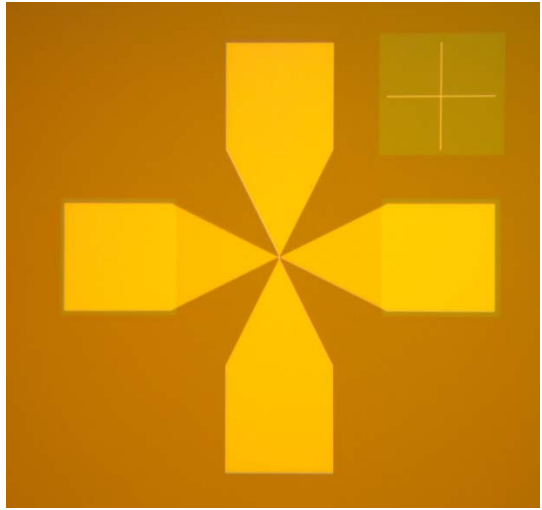

Figure S1: Device structure

- Device fabrication

The Pt/TiO$_2$(30nm)/Pt devices have been implemented in "bone-structure" geometry with active area ~ 1 µm$^2$ (fig. S1). An evaporated Ti/Pt bottom electrode (5nm/25nm) has been patterned by conventional optical lithography technique on a Si/SiO$_2$ substrate (500 µm/200 nm, respectively). Then, a 30 nm TiO$_2$ switching layer has been realized by atomic layer deposition at 200°C using Titanium Isopropoxyde (C$_{12}$H$_{28}$O$_4$Ti) and water as a precursor and reactant, respectively. A Pt/Au electrode (15nm/25nm) was evaporated on top of the TiO$_2$ blanket layer. A rapid annealing of the device was finally applied at 500° C under N$_2$ and N$_2$+O$_2$ atmosphere for 5 min to improve the crystallinity of the TiO$_2$ material.

- Electrical characterization

We used conventional 2-probe technique for the electrical characterization of the device because electrode resistance is much smaller (~of the order 80 Ω) compared to that of the ON-state of the device. Each device was formed by a negative voltage sweep between 0 and -10V (the forming voltage was around -8V) and a current compliance of 300 µA ensured by a transistor connected in series to the $TiO_2$ device. After this forming process, the compliance transistor is removed. Few sweeps between -1.5 V and 1.5 V were applied before obtaining a stable bipolar behavior of the device.

The extended wave form capability of the tool (B1530) was used for the pulse measurement presented in this paper (sampling rate of 10 ns and rise/fall time between 0 and 5V of 80 ns). The Op-amp circuit was realized with a TL072 op amp. During the programming, the devices are disconnected from the OpAmp circuit and programmed using the high precision algorithm setup (B1530 wave form generator and measurement unit). After programming the devices are connected externally to the CMOS chip and the output is measured with an oscilloscope (Agilent 3000 series) while the input pulses are generated with a waveform generator (Agilent 33520). The state of the devices on MAC circuit where set to 15, 30 and 60uA (read current at -200mV).

- Fitting of SET and RESET transitions

We present in this paper the fitting of both SET and RESET transitions. The physical interpretation of the fitting is beyond the scope of this paper and will necessitate more detail analysis and measurements. We are only interesting in the phenomenological behavior and trends of both transitions in order to implement an effective algorithm using the two transitions in an analog way.

We describe the current during the RESET transition with a power law

$$I = A_0 t^{-\beta}$$

, where $\beta$ and $A_0$ are function of voltage V. Figure 2b shows the good agreement between measurement and fitting and figure S2 present the dependence of $A_0$ and $\beta$ with voltage. The current during the SET transition has been fitted with a sigmoidal function:

$$I = A_0 + \frac{A_1 \ln(Bt + 1)}{1 + e^{-\frac{1}{\tau_1}(t - \tau_2)}}$$

Where $A_0$ and $A_1$ are constant ($9 \times 10^{-6}$ A and $7.84 \times 10^{-5}$ A, respectively) fixe to the same value for all the SET transitions and $B$, $\tau_1$ and $\tau_2$ depends on voltage V. We present in figure 2c the fitting of the measurement and in figures S3 a and b the dependence of $B$, $\tau_1$ and $\tau_2$ with voltage. The first stage of the SET transition is well described by an exponential dependence with time (at least before the pseudo saturation that follows a logarithmic dependency). We can also underline the exponential dependence of $\tau_1$ (equivalent to the time for switching) and $\tau_2$ (equivalent to the time for starting switching) with voltage that is consistent with the extrapolation of ns switching time for $TiO_2$ devices.

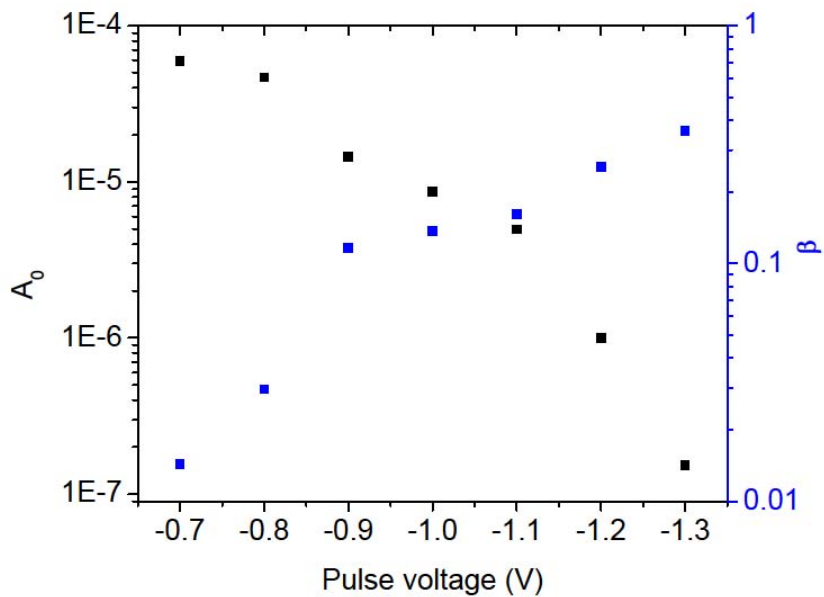

Figure S2: fitting parameters of the RESET transition

**(a)**

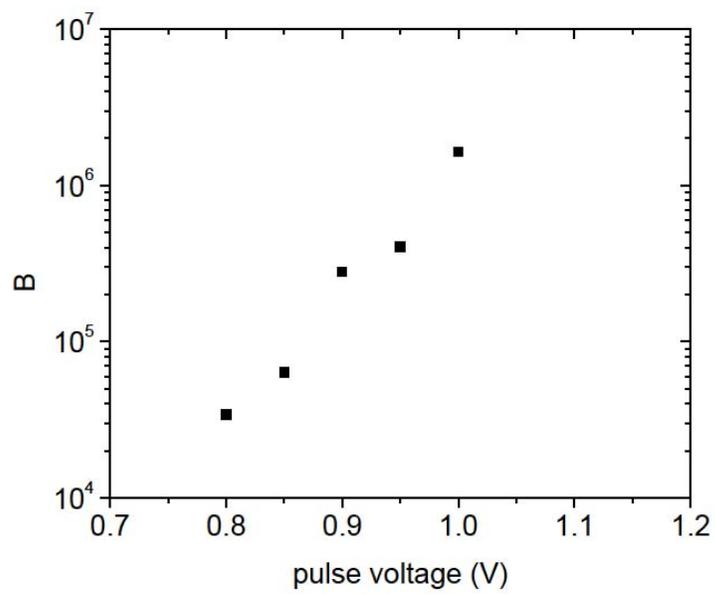

**(b)**

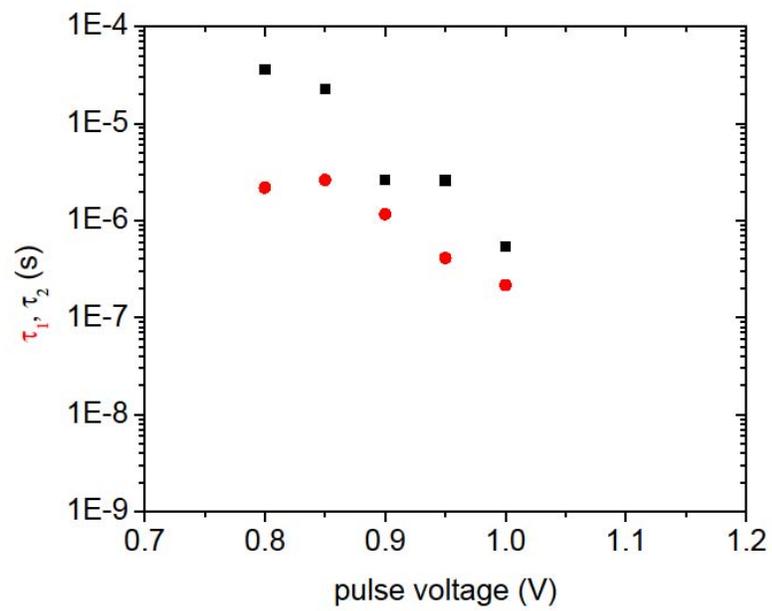

Figure S3 a and b: fitting parameters of the SET transition